\documentclass[a4paper,11pt]{article}
\usepackage{jheppub} 
\usepackage{lineno}
\usepackage{amsmath}
\usepackage{subfigure}
\usepackage{mathrsfs}
\usepackage{color}

\def\bc{\begin{center}}

\def\ec{\end{center}}
\def\be{\begin{eqnarray}}
\def\ee{\end{eqnarray}}

\title{\boldmath Revisit the relationship between spread complexity rate and radial momentum}

\author[a]{Peng-Zhang He}
\affiliation[a]{School of Physics and Astronomy, China West Normal Uniersity, Nanchong 637002, Sichuan, China}

\emailAdd{hepzh@cwnu.edu.cn}

\abstract{This article discusses the relationship between the boundary spread complexity rate and the radial momentum in the bulk within the framework of AdS/CFT. We demonstrate that the radial momentum of a freely falling particle, as measured by a stationary observer in the bulk, is equal to the spread complexity rate of the boundary conformal field theory. This relationship has been discussed in two previous works, but their formulations appear distinct. However, this paper demonstrates their consistency. While earlier studies focused on particles with large mass and massless particles, we show that the relationship \textit{formally} holds for particles of any mass. Additionally, we provide a simple method for obtaining spread complexity from radial momentum using optical geometry.}

\begin{document}
\maketitle
\flushbottom
   
\section{Introduction}
Complexity is an important subject in the study of physics. Recently, a quantity known as Krylov complexity \cite{Parker:2018yvk} has been proposed for studying the Heisenberg evolution of operators. { This concept has attracted much attention and developed rapidly since it was first proposed.}

Krylov complexity, as a measure for studying the properties of operators, has been proposed and is well-defined in any quantum system. Therefore, there has been a considerable amount of research in various quantum many-body systems \cite{Caputa:2022eye,Trigueros:2021rwj,Bhattacharjee:2022ave,Bhattacharjee:2023uwx} and quantum field theories\cite{Dymarsky:2021bjq,Avdoshkin:2022xuw,Adhikari:2022whf,Camargo:2022rnt,Vasli:2023syq,Chattopadhyay:2024pdj,He:2024xjp,He:2024hkw}. In addition, Krylov complexity has various generalizations, such as being extended to open quantum systems \cite{Liu:2022god,Bhattacharya:2023zqt,Mohan:2023btr,Bhattacharya:2024uxx} and being applied to study the evolution of quantum states \cite{Balasubramanian:2022tpr,Caputa:2022yju,Afrasiar:2022efk,Pal:2023yik,Huh:2023jxt}, among others. Krylov complexity has seen numerous purely theoretical developments, such as the use of Lie algebra and coherent state to calculate Krylov complexity \cite{Caputa:2021sib}, as well as the Toda chain method \cite{Dymarsky:2019elm} for computing Krylov complexity, and so on. For more related research, see
 \cite{Ganguli:2024uiq,Fu:2024fdm,Nandy:2024mml,Fan:2024iop,Xu:2024gfm,Caputa:2024sux,Baggioli:2024wbz,Rabinovici:2020ryf,Bhattacharjee:2022vlt,Rabinovici:2021qqt,Erdmenger:2023wjg,Bhattacharjee:2022qjw,Bhattacharya:2022gbz,Bhattacharjee:2022lzy,Camargo:2023eev,Camargo:2024deu,He:2022ryk,Caputa:2024vrn,Hornedal:2022pkc,Bento:2023bjn,Nandy:2023brt,Bhattacharjee:2024yxj,Nandy:2024zcd,Nandy:2024htc,Das:2024zuu,Balasubramanian:2024ghv,Aguilar-Gutierrez:2023nyk,Zhai:2024odw,Zhai:2024tkz,Li:2024ljz,Li:2024kfm,Li:2024iji}.

{ While much research exists on Krylov complexity, its application in holography remains limited.} In this article, we will discuss, within the framework of AdS/CFT \cite{Maldacena:1997re}, the recently proposed relationship between the spread complexity, which is a version of Krylov complexity for quantum state evolution, and the radial momentum of freely moving particles in the bulk. Similar relationships have been discussed as early as in Susskind's articles on holographic complexity \cite{Brown:2017jil}. Although there have been articles \cite{Ageev:2018msv,Ageev:2019fxn,Ageev:2018nye} that followed the original Susskind paper to study momentum/complexity, progress in related research has been slow due to the lack of a clear definition of complexity in field theory. The introduction of Krylov complexity provides a clear definition, allowing us to attempt to find out what the dual of complexity is in gravity.

In article \cite{Caputa:2024sux}, it is proposed that the spread complexity rate in the boundary conformal field theory (CFT) of AdS/CFT is proportional to the radial momentum of a particle in the bulk that is at the asymptotic boundary and has zero velocity at $t = 0$ and is freely falling. This radial momentum needs to be calculated in a coordinate system that computes the proper distance to the black hole, which is why the authors refer to it as the \textit{proper momentum}. The authors illustrated that such a relationship indeed exists in AdS$_3$/CFT$_2$ by using a massive particle. Another article also stated something similar, that the spread complexity rate is equal to the radial momentum measured by a stationary observer of a massless particle. This paper considers the radial momentum of massless particle measured by a stationary observer in gravity and then integrates along the geodesic to see if the calculation result matches the spread complexity in the boundary field theory.

In this article, we will reorganize the derivations of these two papers and demonstrate through comparison that their statements are consistent. Using the method from \cite{Fan:2024iop}, we will verify that for massive particles, it is not necessary to restrict the scaling dimension of the boundary operator to satisfy $\Delta \gg 1$ in order to obtain that the spread complexity rate equals the radial momentum\footnote{For convenience, most of the time we directly refer to the radial momentum observed by the stationary observer as the radial momentum, which the reader should be mindful of.}. Additionally, we will provide a method for easily calculating the spread complexity in the boundary field theory from the radial momentum using the optical metric.

The structure of this paper is as follows: In Section \ref{sec2}, we will provide a brief review of spread complexity. In Section \ref{sec3}, we will compare previous discussions regarding the spread complexity rate and radial momentum, demonstrating that their viewpoints are consistent and deriving the correspondence that holds for particles of any mass. In Section \ref{sec4}, we will utilize optical geometry to present a simple method for calculating spread complexity from radial momentum. Section \ref{sec5} will provide a summary of this paper.

\section{Spread complexity}\label{sec2}
In this section, we will briefly review the concept of spread complexity \cite{Balasubramanian:2022tpr}. For a quantum system described by a time-independent Hamiltonian $H$, the evolution of the state $\left |\psi(t)\right >$ satisfies the Schrödinger equation
\begin{equation}
	i\partial_{t}\left |\psi(t)\right >=H\left |\psi(t)\right >.
\end{equation}
The solution of this equation is 
\begin{equation}
	\left |\psi(t)\right >=e^{-iHt}\left |\psi(0)\right >=\sum_{n=0}^{\infty}\frac{(-it)^{n}}{n!}\left |\psi_{n}\right >,
\end{equation}
where $\left |\psi_{n}\right >=H^{n}\left |\psi(0)\right >$. The evolution of the state $\left |\psi(t)\right >$ over time can always be described by a space spanned by $\{\left |\psi_n\right >\}$. By employing the Lanczos algorithm
\cite{viswanath1994recursion}, we can obtain an orthonormal basis  $\mathcal{K}=\{\left |K_{n}\right >:n=0,1,2,\cdots\}$ for this space, sometimes call Krylov basis. This basis satisfies
\begin{gather}
	\left |K_{0}\right >=\left |\psi(0)\right >,\\
	\left |A_{n+1}\right >=(H-a_{n})\left |K_{n}\right >-b_{n}\left |K_{n-1}\right >,\qquad \left |K_{n}\right >=b^{-1}_{n}\left |A_{n}\right >.\label{2.4}
\end{gather}
Lanczos coefficients $a_{n}$ and $b_{n}$ are defined by
\begin{equation}
	a_{n}=\left <K_{n}\right |H\left |K_{n}\right >,\qquad b_{n}=\left <A_{n}|A_{n}\right >^{1/2},
\end{equation}
and $b_{0}\equiv 0$. The Lanczos algorithm \eqref{2.4} demonstrates that
\begin{equation}
	H\left |K_{n}\right >=a_{n}\left |K_{n}\right >+b_{n+1}\left |K_{n+1}\right >+b_{n}\left |K_{n-1}\right >.
\end{equation}
We can expand \(\left |\psi(t)\right >\) in this basis as follows
\begin{equation}
	\left |\psi(t)\right >=\sum_{n}\psi_{n}(t)\left |K_{n}\right >.
\end{equation}
Spread complexity is defined as
\begin{equation}
	C_{K}(t)=\sum_{n}n|\psi_{n}(t)|^{2}.
\end{equation}
For more detail, it is recommended to refer to \cite{Balasubramanian:2022tpr}.

\section{Spread complexity and radial momentum}\label{sec3}

Recently, in \cite{Caputa:2024sux} and \cite{Fan:2024iop}, it has been independently proposed that the spread complexity on the boundary is equal to the radial momentum in the bulk in the framework of AdS/CFT. However, the formulas conjectured by the two articles seem to be different. In this section, we will restate these two articles and demonstrate that they are consistent.

\subsection{Spread complexity and the radial momentum of massive particles}\label{sec3.1}

In \cite{Caputa:2024sux}, the author considers the evolution of locally excited states by primary operators in holographic CFTs. The excited state is given by
\begin{equation}\label{1}
	\left |\psi(t)\right >=\mathcal{N}e^{-iHt}e^{-\epsilon H}\mathcal{O}(x_0)\left |\psi_\beta\right >,
\end{equation}
where $\mathcal{N}$ is a normalization constant, $\beta=1/T$ is the inverse temperature, $\mathcal{O}(x_0)$ is primary operator with scaling dimension of $\Delta$. Here $x_0$ is the spatial position. $\left |\psi_\beta\right >$ is a thermofield-double state or vacuum state. In \eqref{1}, $\epsilon$ represents an ultraviolet cutoff, which ensures that the energy is finite. This can be seen from the following operation
\begin{equation}
	\begin{aligned}
		\left |\psi\right >&=\sum_{n}\mathcal{N}e^{-iHt}e^{-\epsilon H}\left |n\right >\left <n\right |\mathcal{O}(x_0)\left |\psi_{\beta}\right >=\sum_{n}\mathcal{N}e^{-\epsilon E_{n}}e^{-iHt}\left |n\right >\left <n\right |\mathcal{O}(x_{0})\left |\psi_{\beta}\right >,
	\end{aligned}
\end{equation}
where $\left |n\right >$ is the eigenstate and $E_{n}$ is the eigenvalue of the energy. Due to the presence of $e^{-\epsilon E_n}$, the contributions from states with very high energy are suppressed and $\left |\psi\right >$ has finite energy
\begin{equation}
	E_{\text{CFT}}=\int dx\left <\psi(t)|T_{00}(x)|\psi(t)\right >=\frac{\Delta}{\epsilon}.
\end{equation}
Such states have already been studied in AdS/CFT \cite{Nozaki:2013wia}. The evolution of state \eqref{1} corresponds to placing a static particle at the asymptotic boundary of AdS at the initial time, which undergoes free fall in the bulk.

Consider the simplest case, which is $\beta \rightarrow \infty$, implying that $\left |\psi_\beta\right > = \left |0\right >$. This is a vacuum state on a plane, and its dual description is given by Poincar\'e AdS$_3$
\begin{equation}
	ds^{2}=\frac{-dt^{2}+dx^{2}+dz^{2}}{z^{2}}.
\end{equation}
Here, the AdS radius has been set to be $1$. At $t=0$, inserting the operator $e^{-\epsilon H}\mathcal{O}(x_0)$ corresponds to placing a particle with zero velocity at $x=x_0, z=\epsilon $. For $\Delta\gg1$ and $\epsilon\ll1$, the mass of the particle is $m\simeq \Delta$. For such a particle, it follows a timelike geodesic in the bulk
\begin{equation}
	z(t)=\sqrt{t^2+\epsilon^2}.
\end{equation}
This geodesic can be derived from the particle action 
\begin{equation}
	S=-m\int dt \sqrt{-g_{\mu\nu}\dot{x}^{\mu}\dot{x}^{\nu}}\equiv \int dt\mathcal{L}
\end{equation}
through variation. The radial momentum is:
\begin{equation}\label{3.7}
	P_{z}=\frac{\partial\mathcal{L}}{\partial \dot{z}(t)}=\frac{mt}{\epsilon\sqrt{t^2+\epsilon^2}}.
\end{equation}
However, this result is different from the spread complexity rate of CFT. Specifically, at finite temperature, the spread complexity is \cite{Caputa:2024sux}
\begin{equation}
	C_{K}(t)=\frac{\Delta\beta^{2}}{2\pi^{2}\epsilon^{2}}\sinh^{2}\left (\frac{\pi t}{\beta}\right ),
\end{equation}
and its rate is 
\begin{equation}\label{3.9}
	\dot{C}_{K}(t)=\frac{E_{0}}{\epsilon}\frac{\beta}{2\pi}\sinh\left (\frac{2\pi t}{\beta}\right ).
\end{equation}
For $\beta \rightarrow\infty$,
\begin{equation}\label{3.10}
	C_{K}(t)=\frac{E_{\text{CFT}}t^{2}}{2\epsilon},\qquad\dot{C}_K(t)=\frac{E_{\text{CFT}}}{\epsilon}t.
\end{equation}
This is clearly different from \eqref{3.7}. Reference \cite{Caputa:2024sux} indicates that by letting $z = e^{-\rho}$, then
\begin{equation}\label{3.11}
	P_{\rho}=-\frac{m}{\epsilon}t\Rightarrow \dot{C}_{K}(t)=-\frac{P_{\rho}}{\epsilon}.
\end{equation}
$P_{\rho}$ is called proper momentum. The metric used to calculate this radial momentum is
\begin{equation}
	ds^{2}=d\rho^{2}+e^{2\rho}(-dt^{2}+dx^{2}).
\end{equation}
$\dot{C}_{K}(t)=-\frac{P_{\rho}}{\epsilon}$ is the main conclusion of \cite{Caputa:2024sux}, for a more detailed discussion in various cases, see \cite{Caputa:2024sux}.

\subsection{Spread complexity and the radial momentum of massless particles}\label{sec3.2}

Section \ref{sec3.1} follows the approach of \cite{Caputa:2024sux} and discusses the spread complexity in the context of a massive particle in AdS space freely falling from the asymptotic boundary at zero velocity, where the spread complexity rate is exactly the proper momentum of this particle. Another article \cite{Fan:2024iop} discusses the radial momentum of a massless particle starting from the asymptotic boundary. The author argues that the spread complexity in the boundary field theory is directly equal to the radial momentum observed by the stationary observer\footnote{A stationary observer is referred to as an observer whose worldline coincides with the integral curves of the Killing vector field \cite{Wald:1984rg,liang2023differential}. For any observer, an proper coordinate system can be defined to record events near the worldline. If the spacetime is four-dimensional, this coordinate system can be determined by the observer's worldline and a tetrad along the worldline.} in the bulk for this massless particle
\begin{equation}\label{3.13}
	\frac{dC_{K}}{dt}=P_{r}.
\end{equation}
Their verification is quite straightforward and simple; this section will reorganize the process.

The specific procedure is to determine $P_r$ and then integrate over time $t$ along the geodesic of the massless particle. If the outcome aligns with $C_K(t)$ on the boundary, then the conjecture is satisfied. Consider the general static spherically symmetric metric in a $d$-dimensional spacetime
\begin{equation}\label{3.14}
	ds^{2}=g_{tt}dt^{2}+g_{rr}dr^{2}+\cdots=-h(r)dt^{2}+\frac{dr^2}{f(r)}+\cdots
\end{equation}
We have omitted the coordinates other than time and radial, which does not affect the discussion. These omitted coordinates are collectively denoted as $\{y^\alpha\}$.
Assume the tangent vector of the massless particle's worldline, or wave vector of the photon, is $K^{a}$. Its dual vector satisfies
\begin{equation}
	K_{a}=-\mathcal{E}(dt)_{a}+\mathcal{P}_{r}(dr)_{a},
\end{equation}
and 
\begin{equation}
	0=K^{a}K_{a}=g^{tt}(-\mathcal{E})^{2}+g^{rr}\mathcal{P}_{r}^{2}\Rightarrow \mathcal{P}_{r}=\frac{\mathcal{E}}{\sqrt{h(r)f(r)}},
\end{equation}
where $\mathcal{E}$ is invariant along the geodesic of photon \cite{Townsend:1997ku}. Assume $p$ is a point between the asymptotic boundary and the event horizon of the black hole (if it exists), and there is a stationary observer $G$ at this point. This observer can record the radial momentum of the photon in his/her proper coordinate system. In the problem we are considering, we only need to find such a proper coordinate system, the basis vectors of which are $Z^{a}=\left (\frac{\partial}{\partial\tau}\right )^{a}$ and $N^{a}=\lambda \left (\frac{\partial}{\partial r}\right )^{a}$, and we are not concerned with the other basis vectors. $\tau$ is the proper time of the observer and $\lambda$ is normalization constant. One can check that 
\begin{equation}
	Z^{a}=\frac{1}{\sqrt{-g_{tt}}}\left (\frac{\partial}{\partial t}\right )^{a}, \qquad N^{a}=\frac{1}{\sqrt{g_{rr}}}\left (\frac{\partial}{\partial r}\right )^{a},
\end{equation}
and 
\begin{equation}
	K^{a}=\alpha Z^{a}+\beta N^{a}.
\end{equation}
$\alpha$ is the energy of the photon as measured by the observer, and $\beta$ is the radial momentum of the photon as measured by the observer. Then we have  
\begin{gather}
	\mathscr{E}\equiv \alpha=-Z^{a}K_{a}=\frac{\mathcal{E}}{\sqrt{h(r)}},\label{3.19}\\
	P_{r}=\beta=N^aK_{a}=\frac{\mathcal{E}}{\sqrt{h(r)}}.\label{3.20}
\end{gather}

The energy of CFT is the energy measured by the stationary observer on the asymptotic boundary, taking the metric of the AdS spacetime as
\begin{equation}\label{3.21}
	ds^{2}=-r^{2}dt^{2}+\frac{1}{r^{2}}dr^{2}+\cdots.
\end{equation}
Using \eqref{3.19}, we get the energy of CFT
\begin{equation}
	E_{0}\equiv\mathscr{E}|_{\text{boundary}}=\epsilon \mathcal{E},\qquad \epsilon=\frac{1}{r}.
\end{equation} 
{It is worth noting that the boundary CFT energy $E_{\text{CFT}}$ defined in Section \ref{sec3.1} (which scales as $\Delta/\epsilon$) is mathematically equivalent to the bulk conserved energy $\mathcal{E}$ of the massless particle in Section \ref{sec3.2}. For the remainder of this paper, we will adopt the notation convention consistent with Fan \cite{Fan:2024iop}, where $\mathcal{E}$ represents the bulk conserved energy and the holographic complexity relations are evaluated accordingly.} If the geodesic of the photon is $r=r(t)$ and 
\begin{equation}\label{3.23}
	C_K(t)=\int_{0}^{t}P_{r}(t^{\prime})dt^{\prime}=\int_{0}^{t}\frac{\mathcal{E}}{\sqrt{h(r(t^{\prime}))}}dt^{\prime}
\end{equation}
is satisfied, our task is completed. Due to the existence of a timelike Killing vector field, we can easily rewrite the metric into ingoing coordinates
\begin{equation}
	\begin{aligned}
		ds^{2}&=-h(r)dt^{2}+\frac{dr^{2}}{f(r)}+\cdots\\
		&=-h(r)\left [dt^{2}-\frac{dr^{2}}{f(r)h(r)}\right ]+\cdots\\
		&=-h(r)\left (dt^{2}-dr_{\ast}^{2}\right )+\cdots\\
		&=-h(r)dv^{2}+2\sqrt{\frac{h(r)}{f(r)}}dvdr+\cdots,
	\end{aligned}
\end{equation}
where $v=t+r_{\ast}(r)$ and $r_{\ast}=-\int_{r}^{\infty}\frac{dr^{\prime}}{\sqrt{h(r^{\prime})f(r^{\prime})}}$ is the tortoise coordinate. The geodesic of an ingoing photon is given by $v = \text{constant}$. Without loss of generality, let $v=0$ be the geodesic we consider for the photon. Then we have 
\begin{equation}
	t+r_{\ast}(r)=0\Rightarrow \frac{dt}{dr}=-\frac{1}{\sqrt{h(r)f(r)}}.
\end{equation}
The integral in \eqref{3.23} can be written as
\begin{equation}\label{3.26}
	\int_{0}^{t}P_{r}(t^{\prime})dt^{\prime}=\int_{r(t)}^{\infty}dr^{\prime}\frac{\mathcal{E}}{\sqrt{f(r^{\prime})}h(r^{\prime})}.
\end{equation}
As a simple example, consider the AdS vacuum \eqref{3.21}, we get $h(r)=f(r)=r^{2}$ and 
\begin{equation}
	r_{\ast}=-\frac{1}{r}.
\end{equation}
The geodesic $t+r_{\ast}(r)=0$ becomes
\begin{equation}
	r=\frac{1}{t}.
\end{equation}
Using \eqref{3.23} or \eqref{3.26}, we get
\begin{equation}
		\int_{0}^{t}P_{r}(t^{\prime})dt^{\prime}=\frac{\mathcal{E}t^{2}}{2}=\frac{E_{0}t^{2}}{2\epsilon}.
\end{equation}
Comparing with \eqref{3.10}, this is precisely the spread complexity $C_{K}(t)$ on the boundary. 

\subsection{Discussion on consistency}
In fact, the expressions in \cite{Caputa:2024sux} and \cite{Fan:2024iop} are completely consistent with each other. When calculating the radial momentum of photon, \eqref{3.20} is actually derived from the transformation relationship of vector components between two coordinate systems
\begin{equation}\label{3.30}
	A^{\prime}_{\mu}=\frac{\partial x^{\nu}}{\partial x^{\prime\mu}}A_{\nu},
\end{equation}
where $A_{\mu}=A_{a}(dx^{\mu})_{a},A^{\prime}_{\mu}=A^{a}(dx^{\prime\mu})_{a}$. If we re-express $\mathcal{P}_r$ as $P^{\prime}_r$, according to \eqref{3.20} and \eqref{3.30}, we have
\begin{equation}
	P^{\prime}_{r}=\sqrt{g_{rr}}P_{r}\Rightarrow dr^{\prime}=\frac{dr}{\sqrt{f(r)}}.
\end{equation}
Therefore, $P_r$ is actually the radial momentum in the following metric
\begin{equation}
	ds^{2}=-h(r)dt^{2}+dr^{2}+\cdots.
\end{equation}
Such radial momentum is precisely the proper momentum discussed in \ref{sec3.1} (with a definitional difference of a factor $\epsilon$). Equation \eqref{3.11} differs from equation \eqref{3.13} by a negative sign, which arises from the use of the transformation $z = e^{-\rho}$ instead of $z = e^{\rho}$ when deriving \eqref{3.11}. In principle, both transformations are permissible, and the difference in the negative sign is avoidable.

\subsection{The proper momentum of a particle of any mass is the spread complexity rate}
Reference \cite{Caputa:2024sux} discusses the case of particles with large mass. This section will use the method from Reference \cite{Fan:2024iop} to examine the radial momentum of particles with any mass value. When $t = 0$, the particle is placed on the asymptotic boundary, and our results will reproduce the previously discussed outcomes. This indicates that for massive particles, regardless of their mass, the radial momentum can always determine the spread complexity rate of the boundary CFT. \textit{It is worth pointing out that when the condition $\Delta \gg 1$ fails to hold, the particle picture ceases to be valid, yet the geometrical optics approximation remains applicable. Consequently, we can still utilize the geodesic framework to perform the derivation.}


Here, we only consider the case where the bulk is a three-dimensional AdS and use the method from Section \ref{sec3.2} for the calculation. Consider a three dimensional AdS space in the Poincar\'e  coordinate
\begin{equation}
	ds^{2}=\frac{dz^{2}-dt^{2}+dx^2}{z^{2}}.
\end{equation}
Assume a particle is placed at the position $z = \epsilon, x_i = 0$ in the bulk at time $t = 0$. The geodesic of the particle is \cite{Nozaki:2013wia}
\begin{equation}\label{4.2}
	z(t)=\sqrt{t^{2}+\epsilon^{2}}.
\end{equation}
The momentum of the particle is
\begin{equation}
	P_{a}=-\mathcal{E}(dt)_{a}+\mathcal{P}_{z}(dz)_{a},
\end{equation}
and it satisfies
\begin{equation}\label{3.36}
	g^{ab}P_{a}P_{b}=-m^{2}\Rightarrow \mathcal{P}_{z}=\frac{\sqrt{z^{2}\mathcal{E}^{2}-m^{2}}}{z},
\end{equation}
where $m$ is the mass of the particle. The time direction and radial basis vectors for the static observer are denoted as $Z^{a}=\left (\frac{\partial}{\partial \tau}\right )^{a}$ and $N^{a}=\lambda \left (\frac{\partial}{\partial z}\right )^{a}$, specifically:
\begin{equation}
	Z^{a}=z\left (\frac{\partial}{\partial t}\right )^{a},\qquad N^{a}=z\left (\frac{\partial}{\partial z}\right )^{a}.
\end{equation}
Then we have 
\begin{equation}
	P^{a}=\rho Z^{a}+\delta N^{a},
\end{equation}
where $\rho$ is the energy measured by the observer, and $\delta$ is the radial momentum measured by the observer, and
\begin{equation}\label{4.7}
	\mathscr{E}\equiv \rho =-Z^{a}P_{a}=z\mathcal{E},\qquad P_{z}\equiv N^{a}P_{a}=z\mathcal{P}_{z}=\sqrt{z^{2}\mathcal{E}^{2}-m^{2}}.
\end{equation}
At $t=0$, the energy of the particle observed by the stationary observer at that time and location is
\begin{equation}
	E_{0}=\epsilon\mathcal{E},
\end{equation}
and the radial momentum \eqref{3.36} should be zero, therefore
\begin{equation}
	m=\epsilon\mathcal{E}.
\end{equation}
Substituting back into equation \eqref{4.7} yields
\begin{equation}
	\begin{aligned}
		P_{z}&=\sqrt{z^{2}\left (\frac{m}{\epsilon}\right )^{2}-m^{2}}\\
		&=\sqrt{(t^{2}+\epsilon^{2})\left (\frac{m}{\epsilon}\right )^{2}-m^{2}}\\
		&=\frac{mt}{\epsilon}=\frac{E_{0}t}{\epsilon}.
	\end{aligned}
\end{equation}
This result aligns with Section \ref{sec3.1}, but our discussion here applies to particles of \textit{any mass}, whereas Section \ref{sec3.1} specifically addresses particles with \textit{large mass}. The spread complexity in the dual field theory is:
\begin{equation}\label{4.10}
	C_{K}(t)=\int_{0}^{t}\frac{mt}{\epsilon}dt=\frac{mt^{2}}{2\epsilon}.
\end{equation}
Similarly, the spread complexity rate for the finite-temperature CFT on the boundary can also be derived from the radial momentum of a particle in gravity, as shown in Appendix \ref{a}.

\section{Calculate spread complexity using optical geometry in AdS spacetime}\label{sec4}
In this section, we will present a method to calculate the spread complexity on the asymptotic boundary of AdS without needing to know the particle geodesics. Considering a general static spherically symmetric spacetime \eqref{3.14} and assuming that a photon starts from the asymptotic boundary, we can always consider that there is velocity only in the radial direction in space. Note that a photon travels along null geodesics, and \textit{spatial projection of the null geodesics yields the
light rays, and these are geodesics of the optical metric} \cite{Gibbons:2008rj,Ono:2017pie}. 

Since we are not considering photons with velocities in spatial directions other than the radial direction, we can take the coordinates in these directions as constants. Optical geometry is given by $ds^2 = 0$ \cite{abramowicz1988optical}, resulting in
\begin{equation}
	dt^{2}=\frac{dr^{2}}{h(r)f(r)}.
\end{equation}
Where $dt$ is precisely the affine parameter of the light ray in this Riemannian space of optical geometry. Then we have 
\begin{equation}
	C_{K}(t)=\int_{0}^{t}P_{r}(t^{\prime})dt^{\prime}=-\int_{\infty}^{r(t)}\frac{P_{r}}{\sqrt{h(r^{\prime})f(r^{\prime})}}dr^{\prime}=\int_{r(t)}^{\infty}dr^{\prime}\frac{\mathcal{E}}{\sqrt{f(r^{\prime})}h(r^{\prime})},
\end{equation}
where we have used \eqref{3.20}. The negative sign following the second equality comes from the fact that when $dt > 0$, $dr$ should be less than 0. This result is exactly \eqref{3.26}.

\section{Conclusion}\label{sec5}

In this article, we have discussed the relationship between the spread complexity on the boundary and the radial momentum in the bulk within the framework of AdS/CFT. The results indicate that in AdS/CFT, the spread complexity rate in the boundary conformal field theory is equal to the radial momentum of particles observed by a stationary observer in the bulk. 

Upon re-examining and comparing the correspondence between spread complexity and radial momentum as proposed in references \cite{Caputa:2024sux} and \cite{Fan:2024iop}, we have discovered that they are fundamentally consistent, even though their formulations appear to be different. For the case of massive particles, we have demonstrated that regardless of the mass, the spread complexity on the boundary can always be obtained through the radial momentum in the bulk. We also propose a simple method based on optical geometry that allows for the calculation of spread complexity on the boundary of AdS space without the need to know the particle's geodesic. Our research findings not only deepen our understanding of the AdS/CFT correspondence but also provide new perspectives on the connections between quantum gravity and quantum information.

There are still many questions worth exploring here. If this correspondence truly holds in AdS/CFT, then how should it be proven? AdS/CFT is just one realization of a holographic duality; are there such or similar dualities in other holographic theories? 

\acknowledgments
This work was supported by the National Natural Science Foundation of China (Grants No.1260051567); The Key Project of Sichuan Science and Technology Education Joint Fund (No.25LHJJ0097) and the Sichuan Natural Science Foundation (Grant No.2026NSFSC0746).

\appendix
\section{Radial momentum and the spread complexity of finite-temperature CFT}\label{a}

For a finite-temperature $T=\beta^{-1}$ two-dimensional CFT, the metric of its gravitational dual is \cite{Caputa:2024sux}:
\begin{equation}
	ds^{2}=d\rho^{2}+\frac{4\pi^{2}}{\beta^{2}}\left (-\sinh^{2}\rho dt^{2}+\cosh^{2}\rho dx^{2}\right ),
\end{equation}
where $\rho$ is the radial coordinate. Its geodesics satisfy:
\begin{equation}\label{a2}
	\tanh \rho(t)=\frac{\tanh \rho_{\Lambda}}{\cosh\frac{2\pi t}{\beta}},\qquad \rho_{\Lambda}=\log \frac{\beta}{\pi\epsilon}.
\end{equation}
Assuming the momentum of the particle is
\begin{equation}
	P_{a}=-\mathcal{E}(dt)_{a}+\mathcal{P}_{\rho}(d\rho)_{a},
\end{equation}
and it satisfies
\begin{equation}
	g^{ab}P_{a}P_{b}=-m^{2}\Rightarrow \mathcal{P}_{\rho}=\sqrt{\frac{\beta^{2}}{4\pi^{2}\sinh^{2}\rho}\mathcal{E}^{2}-m^{2}},
\end{equation}
where $m$ is the mass of the particle. The time direction and radial basis vectors for the static observer are denoted as $Z^{a}=\left (\frac{\partial}{\partial \tau}\right )^{a}$ and $N^{a}=\lambda \left (\frac{\partial}{\partial \rho}\right )^{a}$, specifically:
\begin{equation}
	Z^{a}=\frac{\beta}{2\pi \sinh\rho}\left (\frac{\partial}{\partial t}\right )^{a},\qquad N^{a}=\left (\frac{\partial}{\partial \rho}\right )^{a}.
\end{equation}
Then we have 
\begin{equation}
	P^{a}=\kappa Z^{a}+\delta N^{a},
\end{equation}
where $\kappa$ is the energy measured by the observer, and $\delta$ is the radial momentum measured by the observer, and
\begin{equation}\label{a7}
	\mathscr{E}\equiv \kappa =-Z^{a}P_{a}=\frac{\beta}{2\pi \sinh\rho}\mathcal{E},\qquad P_{\rho}\equiv N^{a}P_{a}=\mathcal{P}_{\rho}=\sqrt{\frac{\beta^{2}}{4\pi^{2}\sinh^{2}\rho}\mathcal{E}^{2}-m^{2}}.
\end{equation}
At $t=0$, the energy of the particle observed by the stationary observer at that time and location is
\begin{equation}
	E_{0}=\frac{\beta}{2\pi \sinh\rho_{\Lambda}}\mathcal{E},
\end{equation}
and the radial momentum \eqref{a7} should be zero, therefore
\begin{equation}
	m=\frac{\beta}{2\pi \sinh\rho_{\Lambda}}\mathcal{E}.
\end{equation}
Substituting back into equation \eqref{a7} yields
\begin{equation}
	\begin{aligned}
		P_{\rho}&=\sqrt{\frac{\beta^{2}}{4\pi^{2}\sinh^{2}\rho}\mathcal{E}^{2}-m^{2}}\\
		&=m\sqrt{\frac{\beta^{2}}{4\pi^{2}\sinh^{2}\rho}\left (\frac{2\pi\sinh^{2}\rho_{\Lambda}}{\beta}\right )^{2}-1}\\
		&=m\sqrt{\frac{\sinh^{2}\rho_{\Lambda}}{\sinh^{2}\rho}-1}\\
		&=m\sqrt{\frac{\sinh^{2}\rho_{\Lambda}(1-\tanh^{2}\rho)}{\tanh^{2}\rho}-1}.
	\end{aligned}
\end{equation}
Combining this equation with \eqref{a2} satisfied by the geodesic yields:
\begin{equation}
	\begin{aligned}
		P_{\rho}&=m\sqrt{\cosh^{2}\rho_{\Lambda}\sinh^{2}\frac{2\pi t}{\beta}}\\
		&=\frac{\beta m}{2\pi \epsilon}\sinh\frac{2\pi t}{\beta}+\frac{m\pi\epsilon}{2\beta}\sinh\frac{2\pi t}{\beta}+O(\epsilon^{3}).
	\end{aligned}
\end{equation}
Neglecting the higher-order terms yields:
\begin{equation}
	P_{\rho}=\frac{\beta m}{2\pi \epsilon}\sinh\frac{2\pi t}{\beta}=\frac{\beta E_{0}}{2\pi \epsilon}\sinh\frac{2\pi t}{\beta}.
\end{equation}
This is precisely the spread complexity rate for the CFT on the boundary \eqref{3.9}.

\bibliographystyle{jhep}
\bibliography{references.bib}
\end{document}